\documentclass[11pt,a4paper]{article}
\pdfoutput=1
\usepackage{jheppub}

\newcommand\nn{\nonumber}
\newcommand\fft[2]{\frac{#1}{#2}}
\newcommand\ft[2]{{\textstyle\frac{#1}{#2}}}

\DeclareMathOperator{\sech}{sech}
\DeclareMathOperator{\sgn}{sgn}

\DeclareMathOperator{\Ai}{Ai}
\renewcommand{\Im}{\operatorname{Im}}
\renewcommand{\Re}{\operatorname{Re}}

\begin{document}

\preprint{LCTP-20-20}

\title{The large-$N$ partition function for non-parity-invariant Chern-Simons-matter theories}

\author{James T. Liu}
\emailAdd{jimliu@umich.edu}

\author{and Xiuyuan Zhang}
\emailAdd{xyzhng@umich.edu}

\affiliation{Leinweber Center for Theoretical Physics, Randall Laboratory of Physics\\ The University of Michigan, Ann Arbor, MI 48109-1040}

\abstract{We extend the Fermi gas approach to a class of ABJM-like necklace quiver theories without parity invariance.  The resulting partition function on $S^3$ retains the form of an Airy function, but now includes a phase that scales as $Nk$ in the large-$N$ limit where $k$ is an overall Chern-Simons level.  We demonstrate the presence of this phase both analytically and numerically in the case of a three node quiver.}

\maketitle

\section{Introduction}

One of the remarkable aspects of AdS/CFT duality is the connection that it makes between quantum gravity and quantum field theory.  A key example of this can be seen in recent work connecting AdS black hole microstate counting and supersymmetric indices \cite{Benini:2015eyy,Cabo-Bizet:2018ehj,Choi:2018hmj,Benini:2018ywd}.  More generally, enormous progress on computing supersymmetric partition functions using localization and other powerful techniques has led to new insights on quantum gravity and M-theory.

While AdS/CFT comes in many different flavors, in addition to $\mathcal N=4$ super-Yang-Mills, one of the most widely analyzed setups is that of supersymmetric Chern-Simons-matter (CSM) theories in the context of AdS$_4$/CFT$_3$ duality.  The most familiar case is that of ABJM theory dual to M-theory on AdS$_4\times S^7/\mathbb Z_k$ \cite{Aharony:2008ug}.  However, it is only a single element of a large family of CSM theories exhibiting $N^{3/2}$ scaling of the sphere partition function \cite{Jafferis:2008qz,Imamura:2008nn,Drukker:2010nc,Herzog:2010hf,Martelli:2011qj,Cheon:2011vi,Jafferis:2011zi}.  In a series of remarkable developments, the ABJM partition function was shown to have the structure of an Airy function \cite{Drukker:2011zy,Fuji:2011km,Marino:2011eh}
\begin{equation}
    Z_{\mathrm{ABJM}}\sim\Ai\left(\left(\fft2{\pi^2k}\right)^{-1/3}\left(N-\fft{k}{24}-\fft1{3k}\right)\right),
\end{equation}
where the Chern-Simons levels are given by $k$ and $-k$.

In fact, it has been further conjectured and shown in many examples that the Airy function behavior
\begin{equation}
    Z=C^{-1/3}e^A\Ai\left(C^{-1/3}(N-B)\right)+Z_{\mathrm{np}},
    \label{eq:Airy}
\end{equation}
is universal for parity-invariant $\mathcal N\ge3$ CSM theories with $N^{3/2}$ scaling of the free energy \cite{Marino:2011eh,Mezei:2013gqa,Honda:2014ica,Moriyama:2014gxa,Assel:2015hsa,Moriyama:2015jsa}.  Here $A$, $B$ and $C$ are $N$-independent coefficients that can depend on the Chern-Simons levels $k_a$ as well as the flavored matter content of the theory.  Expansion of the Airy function then immediately gives the large-$N$ asymptotics
\begin{equation}
    F=\log Z\sim-\fft23C^{-1/2}N^{3/2}+BC^{-1/2}N^{1/2}-\fft14\log N+\cdots.
\end{equation}
It is important to keep in mind, however, that this Airy function behavior was originally obtained in parity-conserving theories and is most easily seen in the Fermi gas approach pioneered in \cite{Marino:2011eh}.

In this paper we extend the Fermi gas approach to a set of $\mathcal N=3$ necklace quiver theories without parity invariance.  For such parity-violating theories, the free energy may have an imaginary part, and we find
\begin{equation}
    Z= C^{-1/3}e^{\tilde A}e^{i\fft{D}{2C}N}\Ai\left(C^{-1/3}\left(N-B-\fft{D^2}{4C}\right)\right)+Z_{\mathrm{np}},
\label{eq:Airynew}
\end{equation}
which reduces to (\ref{eq:Airy}) in the parity conserving case where $D=0$.  Here the $N$-independent coefficient $\tilde A$ may in general be complex.  This form of the partition function immediately yields a distinctive signature of parity violating theories in the form of an imaginary term linear in $N$ in the free energy
\begin{equation}
    F\sim-\fft23C^{-1/2}N^{3/2}+i\fft{D}{2C}N+\left(B+\fft{D^2}{4C}\right)C^{-1/2}N^{1/2}-\fft14\log N+\cdots.
\label{eq:Fexp}
\end{equation}
While this result is obtained in the M-theory limit, the imaginary linear-$N$ term scales as $Nk$ where $k$ is the `overall' Chern-Simons level, and can be also be obtained from the genus-zero free energy in the 't~Hooft limit.

The structure of this paper is as follows.  In the next section we review the Fermi gas method as applied to a family of necklace quivers without fundamental matter.  We then compute the partition function in the large-$N$ limit and demonstrate that it reduces to the Airy function form given above.  In section~\ref{sec:three}, we specialize to three node quivers as an example of the general formalism and in particular give explicit expressions for the $B$, $C$ and $D$ parameters in terms of the Chern-Simons levels.  We also provide a numerical confirmation of the imaginary linear-$N$ term as a check on our results.  Finally, we conclude in section~\ref{sec:disc} with a few remarks on parity invariance and on connecting to the 't~Hooft limit.

\section{The Fermi gas approach to necklace quivers}

Before proceeding, we start with a brief review of the Fermi gas approach to $\mathcal N=3$ CSM theories.  The particular models we consider are the $U(N)^r$ necklace quiver theories with $r$ nodes connected by pairs of bi-fundamental chiral superfields \cite{Jafferis:2008qz,Imamura:2008nn}.  Each gauge group has Chern-Simons level $k_a$, which we write as
\begin{equation}
    k_a=n_ak,\qquad\mbox{with}\qquad\sum_{a=1}^rn_a=0.
\end{equation}
The partition function can be written, using localization, as a matrix integral \cite{Kapustin:2009kz}
\begin{equation}
    Z=\fft1{(N!)^r}\int\prod_a\left[\prod_i\fft{d\lambda_i^{(a)}}{2\pi}\fft{\prod_{i<j}\left(2\sinh\left(\fft{\lambda^{(a)}_i-\lambda^{(a)}_j}2\right)\right)^2}{\prod_{i,j}2\cosh\left(\fft{\lambda^{(a)}_i-\lambda^{(a+1)}_j}2\right)}\exp\left(\fft{ikn_a}{4\pi}\sum_i\lambda_i^{(a)\,2}\right)\right].
\label{eq:Znecklace}
\end{equation}
As demonstrated in \cite{Marino:2011eh}, this can be mapped into an equivalent partition function for a one-dimensional non-interacting Fermi gas with a single particle Hamiltonian $\hat H$ that can be obtained from the density matrix $\hat\rho$ whose Wigner transform has the form
\begin{equation}
    \rho_W(q,p)=\fft1{2\cosh\fft{p}2}\star\fft1{2\cosh\fft{p-n_1q}2}\star\fft1{2\cosh\fft{p-(n_1+n_2)q}2}\star\cdots\star\fft1{2\cosh\fft{p-(n_1+\cdots+n_{r-1})q}2}.
\label{eq:rhoW}
\end{equation}
The overall Chern-Simons level $k$ plays the role of Planck's constant, with the identification $\hbar = 2\pi k$.

Given this mapping to a one-dimensional Fermi gas, we then apply standard statistical mechanics techniques to compute the partition function $Z(N)$.  Here it is useful to work with the grand canonical partition function and corresponding grand canonical potential
\begin{equation}
    \Xi(\mu)=e^{J(\mu)}=1+\sum_{N=1}^\infty Z(N)e^{\mu N}.
\end{equation}
The general procedure is now to compute the grand canonical potential $J(\mu)$ from the quantum Hamiltonian and then to obtain the microcanonical partition function $Z(N)$ from $\Xi(\mu)$.  While several different methods can be applied, one step towards computing $J(\mu)$ is to first compute the number of states $n(E)$ below the Fermi energy $E$ as obtained from the quantum Hamiltonian $\hat H$.  The grand canonical potential is then given by
\begin{equation}
    J(\mu)=\int_0^\infty dE\rho(E)\log(1+e^{\mu-E}),
\label{eq:Jfromn}
\end{equation}
where the density of states is $\rho(E)=dn(E)/dE$.  From here we finally obtain the microcanonical partition function through the transform
\begin{equation}
    Z(N)=\fft1{2\pi i}\int d\mu e^{J(\mu)-\mu N}.
\label{eq:ZfromJ}
\end{equation}
Although the Fermi gas picture remains valid at finite $N$, the problem often simplifies in the large-$N$ limit, which corresponds to the thermodynamic limit of the Fermi gas.

As demonstrated in \cite{Marino:2011eh}, for a large class of parity conserving CSM theories, the number function takes the form
\begin{equation}
    n(E)=CE^2+n_0+\mathcal O(Ee^{-E}),\qquad E\gg1.
\end{equation}
Substituting this into (\ref{eq:Jfromn}) results in a grand canonical potential
\begin{equation}
    J(\mu)=\fft13C\mu^3+B\mu+A+\mathcal O(\mu e^{-\mu}),
\label{eq:grand0}
\end{equation}
where $B=n_0+\pi^2C/3$ and $A$ is a $\mu$-independent constant that however depends on $k$.  Application of (\ref{eq:ZfromJ}) then immediately gives the Airy function form, (\ref{eq:Airy}), of the microcanonical partition function $Z(N)$.  In general, the coefficients $A$, $B$ and $C$ will depend on $k$ through the mapping $\hbar=2\pi k$.  For ABJM-like theories, $\hbar C=c_0$ is classical while $\hbar B=b_0+b_1\hbar^2$ receives a single quantum contribution at the perturbative level.  The $A$ coefficient, on the other hand, receives contributions at all orders in the $\hbar$ expansion.

So far, the semiclassical treatment outlined above depends on the Hamiltonian being Hermitian, so that the spectrum is real.  This ensures that the counting of states $n(E)$ is well defined and can be mapped into the problem of obtaining the phase space area for a real Fermi energy $E$.  For a necklace quiver of the form (\ref{eq:Znecklace}), however, the quantum Hamiltonian $\hat H$ is not guaranteed to be Hermitian.  In particular, the Wigner transformed Hamiltonian $H_W(q,p)$ corresponding to the density matrix (\ref{eq:rhoW}) may be complex since the star product
\begin{equation}
    *=\exp\left[\fft{i\hbar}2\left(\overset\leftarrow\partial_q\overset\rightarrow\partial_p-\overset\leftarrow\partial_p\overset\rightarrow\partial_q\right)\right],
\label{eq:star}
\end{equation}
involves $i\hbar$ as the expansion parameter.

Even for the standard ABJM case, this issue of a complex $H_W(q,p)$ can show up since the Wigner transformed density matrix can be written as
\begin{equation}
    \rho_W(q,p)=\fft1{2\cosh\fft{p}2}\star\fft1{2\cosh\fft{q}2},
\label{eq:rW1}
\end{equation}
which is equivalent to the symmetric expression
\begin{equation}
    \rho_W^{(\mathrm{sym})}(q,p)=\fft1{\left(2\cosh\fft{q}2\right)^{1/2}}\star\fft1{2\cosh\fft{p}2}\star\fft1{\left(2\cosh\fft{q}2\right)^{1/2}},
\end{equation}
by conjugation with $\left(2\cosh\fft{q}2\right)^{1/2}$.  Evaluating the star product up to $\mathcal O(\hbar^2)$ for the symmetric form of the density matrix gives the Wigner transformed Hamiltonian
\begin{align}
    H_W^{(\mathrm{sym})}(q,p)&=\log\left(2 \cosh\fft{p}2\right) + \log\left(2 \cosh\fft{q}2\right)\nn\\
    &\quad- \fft{\hbar^2}{192} \left(\sech^2\fft{q}2 \tanh^2\fft{p}2 -\fft12
     \sech^2\fft{p}2 \tanh^2\fft{q}2\right)+\cdots,
\label{eq:HWsym}
\end{align}
which was investigated in \cite{Marino:2011eh}.  However, if we started from (\ref{eq:rW1}), we would instead obtain
\begin{align}
    H_W(q,p)&=\log\left(2\cosh\fft{p}2\right) + \log\left(2 \cosh\fft{q}2\right) + \fft{i\hbar}8 \tanh\fft{p}2 \tanh\fft{q}2\nn\\
    &\quad- \fft{\hbar^2}{192} \left( \sech^2\fft{q}2 \tanh^2\fft{p}2 + 
    \sech^2\fft{p}2 \tanh^2\fft{q}2\right)+\cdots,
\label{eq:HWcmplx}
\end{align}
which has an imaginary term linear in $\hbar$.

As highlighted in \cite{Marino:2011eh}, the number density $n(E)$ can be computed from
\begin{equation}
    n(E)=\fft{\mathrm{Vol}(E)}{2\pi\hbar}+\mathcal O(Ee^{-E}),
\end{equation}
where the phase space area can be obtained from $H_W(q,p)$.  This is straightforward when $H_W(q,p)$ is real, such as in the symmetric case (\ref{eq:HWsym}).  In particular, the starting point is the classical term in the thermodynamic limit
\begin{equation}
    H_0^{(\mathrm{sym})}(q,p)=\fft{|p|}2+\fft{|q|}2,
\end{equation}
which leads to $\mathrm{Vol}_0(E)=8E^2$.  Corrections to this result come from two sources.  The first is the deviation away from the thermodynamic limit
\begin{equation}
    \Delta H_0^{(\mathrm{sym})}(q,p)=\left[\log\left(2\cosh\fft{p}2\right)-\fft{|p|}2\right] +\left[\log\left(2 \cosh\fft{q}2\right)-\fft{|q|}2\right],
\end{equation}
and the second is the quantum correction
\begin{equation}
    \Delta H_q^{(\mathrm{sym})}(q,p)=- \fft{\hbar^2}{192} \left(\sech^2\fft{q}2 \tanh^2\fft{p}2 -\fft12\sech^2\fft{p}2 \tanh^2\fft{q}2\right).
\end{equation}
As demonstrated in \cite{Marino:2011eh}, it is only necessary to work to $\mathcal O(\hbar^2)$ for perturbative corrections to the area.  Combining both sources of corrections then gives \cite{Marino:2011eh}
\begin{equation}
    \mathrm{Vol}(E)=8E^2-\fft{4\pi^2}3+\fft{\hbar^2}{24}+\mathcal O(Ee^{-E}),
\label{eq:VolABJM}
\end{equation}
which leads to the standard result
\begin{equation}
    C=\fft4{\pi\hbar},\qquad B=\fft{2\pi}{3\hbar}+\fft{\hbar}{48\pi},
\end{equation}
where $\hbar=2\pi k$.

Computing the phase space area is less obvious in the case of the complex Wigner transformed Hamiltonian of (\ref{eq:HWcmplx}).  However, we can analytically continue $k\to ik$, which is equivalent to formally taking $i\hbar$ to be real.  Even with this analytic continuation, the $i\hbar$ term in (\ref{eq:HWcmplx}) must be treated with care, as it is not exponentially suppressed in the large $|p|$ and $|q|$ limit.  This means the appropriate starting point in the thermodynamic limit is the semi-classical expression
\begin{equation}
    H_0(q,p)=\fft{|p|}2+\fft{|q|}2+\fft{i\hbar}8\sgn(p)\sgn(q).
\label{eq:H0ABJM}
\end{equation}
The curve $H_0(q,p)=E$ still defines a polygonal region in phase space, however with vertices shifted by $\pm i\hbar/4$.  Because of the shifted vertices, the area of this region is now $\mathrm{Vol}_0(E)=8E^2-\hbar^2/8$.  The complete area again receives two corrections, the first from
\begin{align}
    \Delta H_0(q,p)&=\left[\log\left(2\cosh\fft{p}2\right)-\fft{|p|}2\right] +\left[\log\left(2 \cosh\fft{q}2\right)-\fft{|q|}2\right]\nn\\
    &\quad+\fft{i\hbar}8\left(\tanh\fft{p}2\tanh\fft{q}2-\sgn(p)\sgn(q)\right),
\end{align}
and the second from
\begin{equation}
    \Delta H_q(q,p)=- \fft{\hbar^2}{192} \left(\sech^2\fft{q}2 \tanh^2\fft{p}2 +\sech^2\fft{p}2 \tanh^2\fft{q}2\right).
\end{equation}
While $\Delta H_0(q,p)$ has a term linear in $\hbar$, its contribution to the shifted volume actually drops out since it contributes oppositely in different regions of phase space.  In particular, the contribution in quadrants I and III cancels that in quadrants II and IV of the $(q,p)$ plane.  The result is
\begin{align}
    \mathrm{Vol}(E)&=\mathrm{Vol}_0(E)+\Delta\mathrm{Vol}_0(E)+\Delta\mathrm{Vol}_q(E)\nn\\
    &=\left(8E^2-\fft{\hbar^2}8\right)-\fft{4\pi^2}3+\fft{\hbar^2}6\nn\\
    &=8E^2-\fft{4\pi^2}3+\fft{\hbar^2}{24},
\end{align}
which agrees with the area, (\ref{eq:VolABJM}), obtained from the real Wigner transformed Hamiltonian $H_W^{(\mathrm{sym})}(q,p)$.

\subsection{Necklace quivers without parity invariance}

In the ABJM case, it is perhaps more straightforward to work exclusively with a Hermitian quantum Hamiltonian.  However, necklace quivers in general without parity invariance will necessarily lead to a complex $H_W(q,p)$.  In particular, for the family of necklace quivers with corresponding density matrix (\ref{eq:rhoW}), we can apply the Baker-Campbell-Hausdorff (BCH) formula to lowest order to obtain
\begin{equation}
    H_W(q,p)=\sum_i U_i-\ft12\sum_{i<j}[U_i,U_j]+\cdots,
\label{eq:HWlo}
\end{equation}
where
\begin{equation}
    U_i=\log\left(2\cosh\fft{p-c_iq}2\right),
\end{equation}
with
\begin{equation}
   c_i=n_1+n_2+\cdots+n_{i-1}.
\end{equation}
The parameters $c_i$ are similar to those introduced in \cite{Marino:2011eh}.  However, here the quiver node ordering is retained so that $c_i$ refers directly to the $i$-th node of the quiver.  In particular, since the Chern-Simons levels $k_i=n_ik$ can be of either sign, the $c_i$'s may not necessarily be arranged in numerical order.

Evaluating the commutators using the star product (\ref{eq:star}) gives the Wigner transformed Hamiltonian
\begin{equation}
    H_W(q,p)=\sum_i\log\left(2\cosh\fft{p- c_iq}2\right)+\fft{i\hbar}8\sum_{i<j}( c_i- c_j)\tanh\fft{p- c_iq}2\tanh\fft{p- c_jq}2+H_q(q,p),
\label{eq:HWoih}
\end{equation}
where $H_q(q,p)$ is the part of the quantum Hamiltonian of $\mathcal O(\hbar^2)$ and higher.  It can be obtained by working to second order in the BCH expansion, with the result
\begin{align}
    H_q(q,p)&=-\fft{\hbar^2}{192}\sum_{i<j}( c_i- c_j)^2\left(\sech^2\fft{p- c_iq}2\tanh^2\fft{p- c_jq}2+\sech^2\fft{p- c_jq}2\tanh^2\fft{p- c_iq}2\right)\nn\\
    &\quad-\fft{\hbar^2}{96}\sum_{i<j<k}\bigg(( c_i- c_j)( c_i- c_k)\sech^2\fft{p- c_iq}2\tanh\fft{p- c_jq}2\tanh\fft{p- c_kq}2\nn\\
    &\kern6em-2( c_j- c_i)( c_j- c_k)\sech^2\fft{p- c_jq}2\tanh\fft{p- c_iq}2\tanh\fft{p- c_kq}2\nn\\
    &\kern6em+( c_k- c_i)( c_k- c_j)\sech^2\fft{p- c_kq}2\tanh\fft{p- c_iq}2\tanh\fft{p- c_jq}2\biggr)\nn\\
    &\quad+\mathcal O(i\hbar^3).
\label{eq:HWoh2}
\end{align}
Note that the sums are taken over the node order of the necklace quiver.

Our goal is to calculate the phase space area arising from this Wigner transformed Hamiltonian.  As above, we break the Hamiltonian into three parts, $H_W(q,p)=H_0(q,p)+\Delta H_0(q,p)+\Delta H_q(q,p)$.  Here we have somewhat abused the notation as the `classical' Hamiltonian includes the $\mathcal O(i\hbar)$ contributions in (\ref{eq:HWoih}), while the `quantum' Hamiltonian only includes terms starting at $\mathcal O(\hbar^2)$, as given in (\ref{eq:HWoh2}).  This is consistent with the choice we made above in (\ref{eq:H0ABJM}) for the ABJM case, and is driven by the fact that the $\mathcal O(i\hbar)$ terms are not exponentially suppressed in any region of phase space.

\begin{figure}[t]
\centering
\includegraphics[width=.65\linewidth]{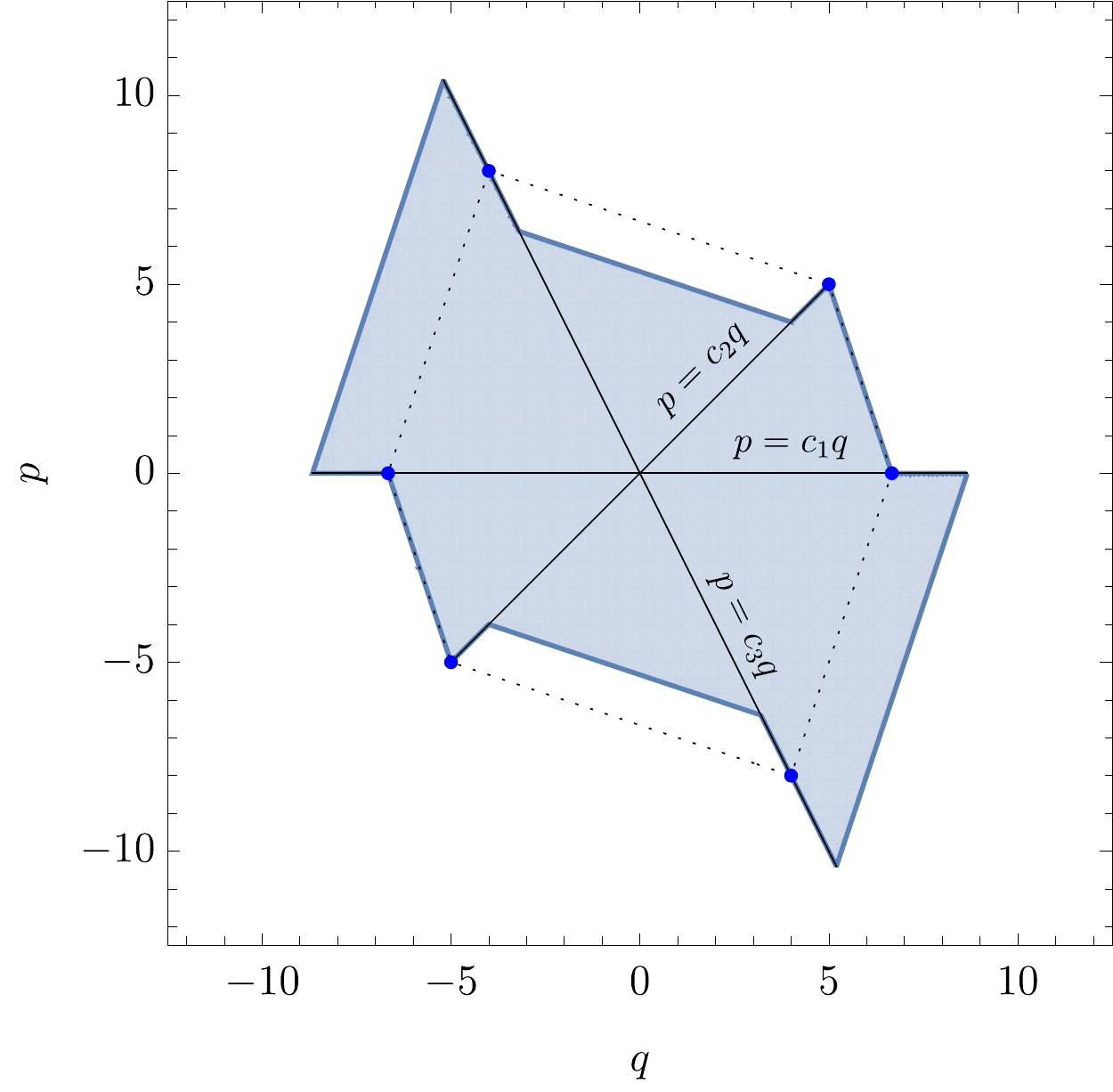}
\caption{An example of the phase space area for a three node quiver computed from (\ref{eq:HWotl}).  Here $n_i=(1,-3,2)$, and we have exaggerated the quantum effect by taking $E=10$ and $i\hbar=4$.  The dotted line indicates the case with $i\hbar=0$.  For this choice of $n_i$'s, we have $c_i=(0,1,-2)$, so the polygon vertex order is $(c_3,c_1,c_2)$, which is different from the quiver node order $(c_1,c_2,c_3)$.\label{fig:poly}}
\end{figure}

We start with the phase space area in the thermodynamic limit.  For $E\gg1$, the `classical' Hamiltonian, (\ref{eq:HWoih}), can be approximated by
\begin{equation}
    H_0(q,p)=\fft12\sum_i|p-c_iq|+\fft{i\hbar}8\sum_{i<j}(c_i-c_j)\sgn(p-c_iq)\sgn(p-c_jq).
\label{eq:HWotl}
\end{equation}
As noted in \cite{Marino:2011eh}, the Fermi surface defined by $H_w(q,p)=E$ is a polygon in phase space, however here with edges shifted by the $\mathcal O(i\hbar)$ terms.  An example is for a three node quiver is shown in Figure~\ref{fig:poly}.  The area can then be calculated by triangulating this polygon.  For an $r$ node quiver, there are generically $2r$ triangles so long as the $c_i$'s are all distinct.  When some of the $c_i$'s are degenerate, the corresponding vertices coincide and some of the triangles degenerate so the polygon will have fewer than $2r$ triangles.  In either case, we only need to take half of the triangles and double the result by reflection symmetry.

There is a bit of a subtlety, however, in that the quiver node order may not correspond to the polygon vertex order.  In particular, the perimeter of the polygon corresponds to the Fermi surface, $H_0(q,p)=E$, and it is triangulated by the rays $p=c_iq$.  Each triangle corresponds to a wedge between two adjacent rays, which can be labeled by the parameters $c_s$ and $c_{\hat s}$ where $c_{\hat s}$ is the next largest parameter following $c_s$
\begin{equation}
    \hat s=\min(i\,|\,c_i>c_s).
\end{equation}
Note that we take $c_{\hat s}$ strictly larger than $c_s$ to avoid degenerate triangles.  This distinction between quiver node order and polygon vertex order can be seen in Figure.~\ref{fig:poly}.

We now consider the triangle bounded by the rays $p=c_sq$ and $p=c_{\hat s}q$.  The outer edge of this triangle is given by $H_0(q,p)=E$ where $p$ lies within the range $c_sq<p<c_{\hat s}q$.  This fixes the absolute values and the signs in (\ref{eq:HWotl}), and the result is that the outside edge is given by
\begin{equation}
    \sum_i|p-c_iq|=2E+\fft{i\hbar}4f_s,
\end{equation}
where
\begin{equation}
    f_s=\sum_i|c_i-c_s|\left(\sum_{j<i}\sgn(c_j-c_s-\epsilon)+\sum_{j>i}\sgn(c_s-c_j+\epsilon)\right).
\label{eq:fs}
\end{equation}
Here $\epsilon\to0^+$ enforces the proper sign for when one or more of the $c_j$'s take the same value as $c_s$.  Note that consistency of the piecewise linear Fermi surface demands $f_s(\epsilon\to0^+)=f_{\hat s}(\epsilon\to0^-)$, so that the shift calculated from vertex $s$ of the triangle matches that calculated from vertex $\hat s$.  Adding up the triangle areas (and doubling the result since each triangle is accompanied by its reflection) then gives
\begin{align}
    \mathrm{Vol}_0(E)&=\sum_s\fft{4|c_{\tilde s}-c_s|}{\sum_j|c_j-c_s|\sum_j|c_j-c_{\tilde s}|}\left(E+\fft{i\hbar}8f_s\right)^2\nn\\
    &=\alpha E^2+i\beta\hbar E+\gamma_0\hbar^2.
\label{eq:Vol0}
\end{align}
Note that, to avoid counting degenerate triangles, the sum over $s$ should be restricted so that each unique value of $c_s$ is to be included only once.  Alternatively, we can sum over all values of $s=1,\ldots,r$ if we define $c_{\hat s}$ to be next in the numerically sorted list including all $c_i$'s even if they are numerically identical.  In this case, degenerate triangles have $c_{\hat s}=c_s$ and hence will not contribute to the area.  The leading $\mathcal O(E^2)$ term in (\ref{eq:Vol0}) is identical to that obtained in \cite{Marino:2011eh}, while the $i\beta\hbar E$ term vanishes in parity conserving theories with $f_s=0$.

The $\gamma_0\hbar^2$ term in (\ref{eq:Vol0}) is not the complete picture, as the $\mathcal O(\hbar^2)$ quantum Hamiltonian $H_q(q,p)$ will also contribute at the same order.  However, the $i\beta\hbar E$ term gives the full contribution at $\mathcal O(i\hbar)$.  This is because the $\mathcal O(i\hbar)$ term in the difference $\Delta H_0(q,p)=H_W(q,p)-H_0(q,p)$ is odd around each vertex of the polygon so its contribution to the area vanishes.  At the same time, the difference between the first terms in (\ref{eq:HWoih}) and (\ref{eq:HWotl}) survives, and gives the standard shift
\begin{equation}
    \Delta\mathrm{Vol}_0(E)=-\fft{2\pi^2}3\sum_s\fft1{\sum_j|c_j-c_s|}.
\label{eq:dVol0}
\end{equation}

We now turn to the corrections arising from the quantum Hamiltonian, (\ref{eq:HWoh2}), which can be rewritten as
\begin{equation}
    H_q(q,p)=-\fft{\hbar^2}{192}\sum_i\sech^2\fft{p-c_iq}2\left(\sum_{j,k}(1-3\delta_i^{j,k})(c_i-c_j)(c_i-c_k)\tanh\fft{p-c_jq}2\tanh\fft{p-c_kq}2\right),
\end{equation}
where the sum over $j$ and $k$ is unrestricted, and $\delta_i^{j,k}$ vanishes unless $i$ lies between $j$ and $k$
\begin{equation}
    \delta_i^{j,k}=\begin{cases}1,&j<i<k\mbox{ or }k<i<j;\\0&\mbox{otherwise}.\end{cases}
\end{equation}
As highlighted in \cite{Marino:2011eh}, $H_q(q,p)$ vanishes exponentially away from the vertices of the polygon.  Integrating the quantum correction around each vertex then gives
\begin{equation}
    \Delta\mathrm{Vol}_q(E)=\fft{\hbar^2}6\sum_{i<j}| c_i- c_j|-\fft{\hbar^2}2\sum_s\fft{\left(\sum_{j<s}| c_s- c_j|\right)\left(\sum_{j>s}| c_s- c_j|\right)}{\sum_j| c_s- c_j|}.
\label{eq:dVolq}
\end{equation}
Finally, adding up the contributions from (\ref{eq:Vol0}), (\ref{eq:dVol0}) and (\ref{eq:dVolq}) gives the area of the Fermi surface
\begin{equation}
    \mathrm{Vol}(E)=\alpha E^2+i\beta\hbar E+\gamma\hbar^2+\delta+\mathcal O(Ee^{-E}),
\end{equation}
where
\begin{align}
    \alpha&=\sum_s\fft{4|c_{\tilde s}-c_s|}{\sum_j|c_j-c_s|\sum_j|c_j-c_{\tilde s}|},\nn\\
    \beta&=\sum_s\fft{|c_{\tilde s}-c_s|f_s}{\sum_j|c_j-c_s|\sum_j|c_j-c_{\tilde s}|},\nn\\
    \gamma&=-\fft1{16}\sum_s\fft{|c_{\tilde s}-c_s|f_s^2}{\sum_j|c_j-c_s|\sum_j|c_j-c_{\tilde s}|}+\fft16\sum_{i<j}| c_i- c_j|\nn\\
    &\qquad-\fft12\sum_s\fft{\left(\sum_{j<s}| c_s- c_j|\right)\left(\sum_{j>s}| c_s- c_j|\right)}{\sum_j| c_s- c_j|},\nn\\
    \delta&=-\fft{2\pi^2}3\sum_s\fft1{\sum_j|c_j-c_s|}.
\label{eq:coefs}
\end{align}

It is now a straightforward exercise to compute the number function $n(E)$ and the grand potential $J(\mu)$ up to non-perturbative corrections.  The result is
\begin{equation}
    J(\mu)=\fft13C\mu^3+\fft{iD}2\mu^2+B\mu+A+\mathcal O(\mu e^{-\mu}),
\label{eq:JmuA}
\end{equation}
where
\begin{equation}
    C=\fft\alpha{2\pi\hbar},\qquad D=\fft\beta{2\pi},\qquad B=\fft{\gamma\hbar}{2\pi}+\fft\delta{2\pi\hbar}+\fft{\pi\alpha}{6\hbar}.
\label{eq:BCD}
\end{equation}
(Recall that $\hbar=2\pi k$.)  The new feature here compared with (\ref{eq:grand0}) is the presence of an imaginary term quadratic in $\mu$.  Nevertheless, the transform (\ref{eq:ZfromJ}) can still be performed by making a constant shift in $\mu$ so as to eliminate the quadratic term.  The resulting partition function then takes the Airy function form (\ref{eq:Airynew}), which we repeat here for convenience
\begin{equation}
    Z= C^{-1/3}e^{\tilde A}e^{i\fft{D}{2C}N}\Ai\left(C^{-1/3}\left(N-B-\fft{D^2}{4C}\right)\right)+Z_{np}.
\label{eq:Zairy2}
\end{equation}
The $N$ independent coefficient $\tilde A$ is related to $A$ in (\ref{eq:JmuA}), but is now complex as it takes on an imaginary component that arises from shifting $\mu$.

\section{The three-node quiver}
\label{sec:three}

As an example, we consider a three-node necklace quiver with levels $k_1$, $k_2$ and $k_3$, or equivalently $n_1$, $n_2$ and $n_3$.  Since the three $n_i$'s sum to zero (and none of them vanish), one of them must have the opposite sign as the other two.  To be specific, we can take $n_1$ and $n_2$ positive and $n_3=-(n_1+n_2)$ negative.  The opposite sign case can be obtained by a parity transformation.  Note also that the three-node quiver is always parity violating, so we should not be surprised to see the imaginary $D$ term show up in (\ref{eq:Zairy2}).

We now take
\begin{equation}
    c_i=(0,n_1,n_1+n_2),
\end{equation}
and note that in this case they are already in numerical order so we may substitute $c_{\hat s}=c_{s+1}$ in (\ref{eq:coefs}).  At the purely classical level, and in the thermodynamic limit, the Fermi surface is an irregular hexagon with three pairs of parallel opposing sides.  The phase space area is then $\alpha E^2$ to lowest order where
\begin{equation}
    \alpha=\frac{8(|n_1 n_2|+|n_1 n_3|+|n_2 n_3|)}{(|n_1|+|n_2|)(|n_1|+|n_3|)(|n_2|+|n_3|)}.
\end{equation}
This picks up a classical correction $\delta$ away from the thermodynamic limit where
\begin{equation}
    \delta =-\frac{2\pi ^2}{3} \left(\frac{1}{|n_1|+|n_2|}+\frac{1}{|n_1|+|n_3|}+\frac{1}{|n_2|+|n_3|}\right).
\end{equation}

We now turn to the quantum corrections.  The $i\beta\hbar E$ correction vanishes in parity conserving theories like the ABJM model.  However, the three node quiver is always parity violating, and we find
\begin{equation}
    \beta=-\frac{2n_1 n_2 n_3}{(|n_1|+|n_2|)(|n_1|+|n_3|)(|n_2|+|n_3|)}.
\end{equation}
Note that this flips sign under $n_i\to-n_i$.  As discussed in the previous section, $\mathcal O(i\hbar)$ terms modify the hexagonal Fermi surface by either stretching (for $f_s>0$) or squeezing (for $f_s<0$) pairs of opposite sides.

The $\mathcal O(\hbar^2)$ term, $\gamma$, receives contributions from two sources, $\gamma=\gamma_0+\gamma_1$, where
\begin{equation}
    \gamma_0=-\frac{(|n_1n_2|+|n_2n_3|+|n_3n_1|)(n_1^2+n_2^2+n_3^2)-|n_1n_2 n_3|(|n_1|+|n_2|+|n_3|)}{4(|n_1|+|n_2|)(|n_1|+|n_3|)(|n_2|+|n_3|)},
\end{equation}
arises from the area of the Fermi surface in the thermodynamic limit, (\ref{eq:Vol0}), and
\begin{equation}
    \gamma_1=\frac{1}{6}(|n_1|+|n_2|+|n_3|)-\frac{2|n_1 n_2 n_3|}{(|n_1|+|n_2|+|n_3|)^2},
\end{equation}
is the correction, (\ref{eq:dVolq}), from the quantum Hamiltonian.  Adding these contributions together gives
\begin{equation}
    \gamma=\fft{(|n_1|+|n_2|+|n_3|)^2(n_1^2+n_2^2+n_3^2)}{48(|n_1|+|n_2|)(|n_1|+|n_3|)(|n_2|+|n_3|)},
\end{equation}
for the complete $\mathcal O(\hbar^2)$ correction.

Finally, we can compute the coefficients (\ref{eq:BCD}) that go into the Airy function, (\ref{eq:Zairy2})
\begin{align}
    C&=\fft2{\pi^2}\fft{|k_1k_2|+|k_2k_3|+|k_3k_1|}{(|k_1|+|k_2|)(|k_2|+|k_3|)(|k_3|+k_1|)},\nn\\
    D&=-\fft1\pi\fft{k_1k_2k_3}{(|k_1|+|k_2|)(|k_2|+|k_3|)(|k_3|+k_1|)},\nn\\
    B&=\fft{(|k_1|+|k_2|+|k_3|)^2(k_1^2+k_2^2+k_3^2)}{48(|k_1|+|k_2|)(|k_2|+|k_3|)(|k_3|+k_1|)}+\fft{(|k_1k_2|+|k_2k_3|+|k_3k_1|)-(k_1^2+k_2^2+k_3^2)}{6(|k_1|+|k_2|)(|k_2|+|k_3|)(|k_3|+k_1|)}.
\label{eq:BCD3}
\end{align}
Note that these can be written in terms of the actual Chern-Simons levels $k_a$.  Moreover, the $B$ coefficient has two contributions, the first scaling as $\sim k$ and the second scaling as $\sim1/k$.  This is similar to the ABJM case, where $B_{\mathrm{ABJM}}=k/24+1/3k$.  The $C$ coefficient was first obtained implicitly in \cite{Herzog:2010hf} where the free energy was calculated at leading order and more directly in \cite{Marino:2011eh}.  Here we have extended the latter calculation to include the higher order coefficients $B$ and $D$.

\subsection{Numerical examination of the free energy}

The $D$ coefficient originates from the $\mathcal O(i\hbar)$ correction and vanishes in parity-conserving cases such as ABJM theory.  However, the three-node quiver is always parity violating, so here we have $D\ne0$.  Expanding the Airy function in (\ref{eq:Zairy2}) then gives the large-$N$ but fixed $k$ expression for the free energy
\begin{equation}
    F\sim-\fft23C^{-1/2}N^{3/2}+i\fft{D}{2C}N+\left(B+\fft{D^2}{4C}\right)C^{-1/2}N^{1/2}-\fft14\log N+\cdots.
\end{equation}
This indicates that
\begin{equation}
    \Im F\sim \fft{D}{2C}N=-\fft{\pi}4\fft{k_1k_2k_3}{|k_1k_2|+|k_2k_3|+|k_3k_1|}N+\cdots.
\end{equation}
While this scales as $\mathcal O(N)$, which is subdominant to the $N^{3/2}$ scaling of $\Re F$, it is easy to see numerically as it is the leading order contribution to $\Im F$.

We have examined the free energy numerically in order to highlight its imaginary component.  To match the Fermi gas picture, we numerically solve the saddle point equation for the three-node quiver for fixed small values of $k$.  In particular, we restrict to quivers with $k_1=k_2$, so that the Chern-Simons levels can be written as $k_a=(k,k,-2k)$.  In this case, the expression (\ref{eq:BCD3}) for the parameters $B$, $C$ and $D$ reduce to
\begin{equation}
    C=\fft5{9\pi^2k},\qquad D=\fft1{9\pi},\qquad B=\fft{k}9-\fft1{108k},
\end{equation}
so that
\begin{equation}
    F\sim-\fft{2\pi}{\sqrt5}N^{3/2}k^{1/2}+\fft{i\pi}{10}Nk+\fft{7\pi}{20\sqrt5}\left(1-\fft5{63k^2}\right)N^{1/2}k^{3/2}-\fft14\log N+\cdots.
\label{eq:fgans}
\end{equation}

\begin{figure}[t]
\centering
\includegraphics[width=.65\textwidth]{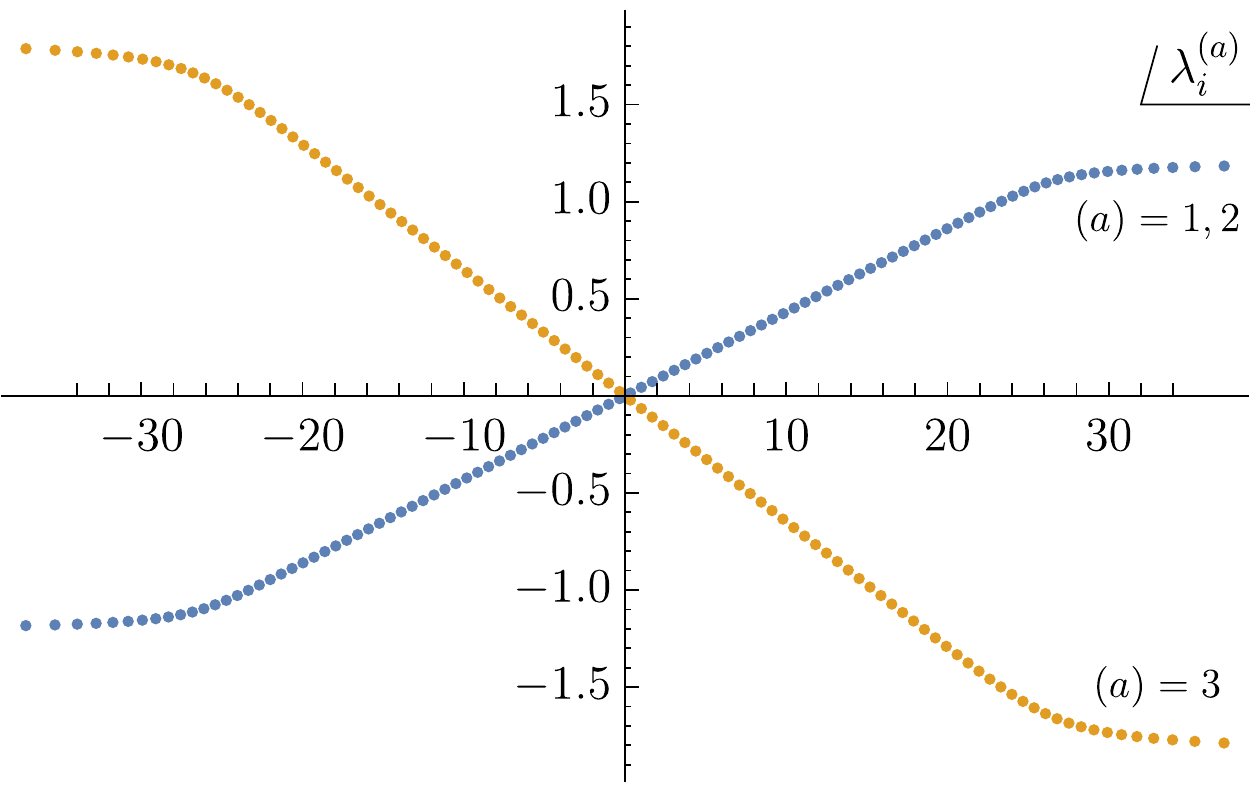}
\caption{The numerical solution to the saddle point equations for $N=100$ and $k_a=(1,1,-2)$.  The first two sets of eigenvalues, $\lambda_i^{(1)}$ and $\lambda_i^{(2)}$ coincide since they have identical Chern-Simons levels.\label{fig:11-2}}
\end{figure}

The numerical procedure is implemented in Mathematica using the built-in FindRoot procedure.  For a given value of $N$ and $k$, FindRoot is called twice, first with WorkingPrecision set to MachinePrecision and subsequently, to refine the solution, set to 50.  The FindRoot procedure is sensitive to the initial trial configuration, which we take to be linear with uniform density for the first time the procedure is run.  Subsequent solutions are obtained using the previous eigenvalue distribution, with the real component scaled appropriately by $\sqrt{N/k}$ when either $N$ or $k$ changes.  All solutions are checked for proper convergence.  An example for $N=100$ and $k_a=(1,1,-2)$ is shown in Fig.~\ref{fig:11-2}.

Once the numerical eigenvalues are obtained, the free energy is computed by evaluating the saddle point action and the Gaussian determinant
\begin{equation}
    F=S_{\mathrm{eff}}(\lambda_i^{(a)})-\fft12\log\det\left(\fft{\partial^2S_{\mathrm{eff}}}{\partial\lambda_i^{(a)}\partial\lambda_j^{(b)}}\right)-\fft{Nr}2\log2\pi+\cdots,
\label{eq:spa}
\end{equation}
where $S_{\mathrm{eff}}(\lambda_i^{(a)})$ is obtained from (\ref{eq:Znecklace})
\begin{align}
    S_{\mathrm{eff}}(\lambda_i^{(a)})&=\sum_a\left[\fft{ikn_a}{4\pi}\sum_i\lambda_i^{(a)\,2}+\sum_{i<j}\log\left(4\sinh^2\left(\fft{\lambda_i^{(a)}-\lambda_j^{(a)}}2\right)\right)\right.\nn\\
    &\kern12em\left.-\sum_{i,j}\log\left(2\cosh\left(\fft{\lambda_i^{(a)}-\lambda_j^{(a+1)}}2\right)\right)\right].
\label{eq:Seff}
\end{align}
The final factor in (\ref{eq:spa}) is a combination of $(2\pi)^{Nr/2}$ from the Gaussian integrals and $1/(2\pi)^{Nr}$ from the integration measure of (\ref{eq:Znecklace}).  Since we are interested in the imaginary component of $F$, care must be taken when evaluating the logs in the effective saddle point action.  For both $\log\sinh(z)$ and $\log\cosh(z)$, we place the branch cuts along the imaginary axis connecting adjacent pairs of zeros.  Evaluating the log of the determinant numerically also introduces an $n\pi i$ ambiguity in the free energy.  This is resolved empirically on a case by case basis when scanning over $N$ by adding multiples of $\pi i$ when appropriate so that $\Im F$ is a smooth function of $N$.  Note, however, that this leaves an overall $N$-independent multiple of $\pi i$ ambiguity in the free energy for each distinct value of $k$.

We numerically explored the free energy in the large-$N$ limit while holding $k_a=(k,k,-2k)$ fixed.  In particular, we scanned over $N$ from $140$ to $300$ for values of $k$ in the range of $1$ to $10$.  For each value of $k$, we performed independent fits for the real and imaginary components of the free energy.  For $\Re F$, we find, for example when $k=1$, the numerical fit
\begin{align}
    \left.\Re F(N,k=1)\right|_{\mathrm{num.}}&=-2.8099N^{3/2}+(3.0000-\ft32\log2\pi)N-0.1611N^{1/2}-0.2309\log N\nn\\
    &\quad-0.5572+0.0313N^{-1/2}+0.0553N^{-1},
\end{align}
where we are not particularly interested in the terms beyond $\log N$ but include them for fitting purposes.  This numerical result can be compared with the Fermi gas solution, (\ref{eq:fgans}), which reads for $k=1$
\begin{equation}
    \left.\Re F(N,k=1)\right|_{\mathrm{Fermi~gas}}=-2.80993N^{3/2}+0.45271N^{1/2}-0.25000\log N+\cdots.
\end{equation}
This demonstrates very good agreement of the leading order $N^{3/2}$ term, but shows discrepancies at the subleading order.  In particular, there is a curious linear-$N$ term in the numerical solution that is not present in the analytic result.

The linear-$N$ behavior was also seen in other numerical investigations such as that of \cite{Liu:2019tuk} (which however pertained to a theory with $N^{5/3}$ scaling of the free energy).  The factor $1-(1/2)\log2\pi$ for each node in the quiver arises as an artifact of terminating the numerical large-$N$ but fixed-$k$ expansion at the Gaussian determinant, as can be seen, for example, in a careful treatment of $1/N$ effects in the leading saddle point expansion of the partition function.  In particular, the effective action, (\ref{eq:Seff}), is log divergent as two eigenvalues on the same node approach each other.  For a single cut saddle point solution, there are $N-1$ adjacent pairs of eigenvalues at each node, each giving rise to a $\log N$ term, for a total of $Nr\log N$ in the saddle point evaluation of $S_{\mathrm{eff}}$.  This log divergence is canceled by a similar factor arising from the Gaussian determinant.  However, the cancellation is not complete, as the constants pertaining to the logs in the `classical' and `one-loop' terms differ by one unit for each node in the quiver.  Combining this factor of $Nr$ with the last term in (\ref{eq:spa}) then gives the $Nr(1-(1/2)\log2\pi)$ factor observed numerically.

Since the numerical leading order $N^{3/2}$ term matches well with the analytic result, we can subtract it as well as the linear-$N$ term from the numerical data and fit the remainder.  We then find numerically
\begin{align}
    \Re F(N,k)&=-\fft{2\pi}{\sqrt5}N^{3/2}k^{1/2}+3\left(1-\fft12\log2\pi\right)N+f_1(k)N^{1/2}+f_2(k)\log N\nn\\
    &\quad+f_3(k)+f_4(k)N^{-1/2}+f_5(k)N^{-1},
\label{eq:Fnum}
\end{align}
where the relevant coefficients $f_1(k)$ and $f_2(k)$ are shown in Table~\ref{tbl:coefs}.  Our first observation is that while $f_2(k)$, the coefficient of $\log N$, is close to the expected $-1/4$, it nevertheless is not a good match and moreover depends non-trivially on $k$.  We are uncertain where this discrepancy originates, but suspect that it is due to a combination of terminating the numerical approximation, (\ref{eq:spa}), at the Gaussian determinant and working with a limited range of $N$.  As for the second point, note in particular that for the numerical data the maximum value of $N=300$ is unchanged even as $k$ is increased.  This is the likely origin of the systematic dependence of $f_2(k)$ on $k$.

\begin{table}[t]
\centering
\begin{tabular}{l|r|r|r}
$k$&$f_1(k)$&$f_2(k)$&$g_1(k)$\\
\hline
$1$&$-0.16078$&$-0.23378$&$0.31416$\\
$2$&$0.92941$&$-0.24006$&$0.62832$\\
$3$&$2.17836$&$-0.24188$&$0.94248$\\
$4$&$3.60757$&$-0.24262$&$1.25664$\\
$5$&$5.20591$&$-0.24318$&$1.57080$\\
$6$&$6.96061$&$-0.24440$&$1.88496$\\
$7$&$8.86057$&$-0.24764$&$2.19911$\\
$8$&$10.89644$&$-0.25489$&$2.51327$\\
$9$&$13.06037$&$-0.26882$&$2.82743$\\
$10$&$15.34571$&$-0.29263$&$3.14158$
\end{tabular}
\caption{The coefficients $f_1(k)$ and $f_2(k)$ for the real component of the numerically determined free energy, (\ref{eq:Fnum}), and $g_1(k)$ for the imaginary component, (\ref{eq:ImF}), for $k=1,\ldots,10$.  The fit is obtained from data for $N=140$ to $300$ in steps of $20$.\label{tbl:coefs}}
\end{table}

The behavior of $f_1(k)$, the coefficient of $N^{1/2}$, is also somewhat curious, as can be seen by comparing it with the Fermi gas result, (\ref{eq:fgans}), which has the form
\begin{equation}
    \left.f_1(k)\right|_{\mathrm{Fermi~gas}}=\fft{7\pi}{20\sqrt5}k^{3/2}-\fft\pi{36\sqrt5}k^{-1/2}=0.49174k^{3/2}-0.03903k^{-1/2}.
\end{equation}
Numerically, we find that $f_1(k)$ does appear to be the sum of two power laws, however with coefficients
\begin{equation}
    \left.f_1(k)\right|_{\mathrm{num.}}=0.49178k^{3/2}-0.65292k^{-1/2}.
\end{equation}
While the $N^{1/2}k^{3/2}$ coefficient agrees well with the analytic result, the $N^{1/2}k^{-1/2}$ coefficient is not at all close.  We should emphasize, however, that we do not view this discrepancy as a failure of either the Fermi gas model or the numerical work.  Instead, as in the behavior of the linear $N$ and $\log N$ terms, we expect the analytic result, (\ref{eq:fgans}), to be valid in the fixed $k$, large $N$ limit, while the corresponding numerical approximation, (\ref{eq:spa}), is missing additional contributions beyond the order of the Gaussian determinant.

In the above, we have looked at the real part of the free energy.  However, for parity-violating quivers such as the three node case, the free energy is complex, and we expect from (\ref{eq:fgans})
\begin{equation}
    \left.\Im F(N,k)\right|_{\mathrm{Fermi~gas}}=\fft\pi{10}Nk+\cdots.
\label{eq:ImF}
\end{equation}
This can be examined numerically, provided the branch issues are treated properly.  For the imaginary component, we fit
\begin{equation}
    \left.\Im F(N,k)\right|_{\mathrm{num.}}=g_1(k)N+g_2(k)+g_3(k)N^{-1/2}+g_4(k)N^{-1},
\end{equation}
where we are mainly interested in the linear-$N$ coefficient $g_1(k)$.  The numerical data is shown in Table~\ref{tbl:coefs}, and indeed matches the linear behavior (\ref{eq:ImF}) with coefficient $\pi/10$ quite well.

In general, the numerical results match the analytic expression for the free energy, (\ref{eq:fgans}).  However, the limitations of working with a fixed set of numerical data and relying on the saddle point approximation, (\ref{eq:spa}), is also apparent.  The noticeable differences between the numerical and analytic results is the addition of a real term linear in $N$ and a mismatch of the $N^{1/2}k^{-1/2}$ coefficient.  In addition, the numerically determined coefficient of $\log N$ is not reliably $-1/4$, in contrast with similar numerical investigations of other models where the $\log N$ coefficient was obtained with reasonable significance \cite{Liu:2017vll,Liu:2018bac,PandoZayas:2019hdb,Liu:2019tuk}.

\section{Discussion}
\label{sec:disc}

The Fermi gas approach has led to a greatly enhanced understanding of ABJM like theories.  However, most of the investigations have been restricted to parity conserving models.  In this paper, we have demonstrated that the Fermi gas picture is easily extended to cover models without parity, where the Hamiltonian of the equivalent Fermi gas system is non-Hermitian.  Our approach is to analytically continue in $i\hbar$ so that intermediate quantities such as the phase space area can be treated formally as real quantities.  Only at the end do we restore $\hbar$ to be real.  This analytic continuation leads to the sphere partition function (\ref{eq:Zairy2}), which retains the form of an Airy function, but with an additional parameter $D$, or equivalently $\beta$, related to a parity violating phase.

As we have seen, one signature of a parity violating theory is the presence of an imaginary component of the free energy that is linear in $N$, with $\Im F\sim(D/2C)N$.  Applying parity to the original quiver is equivalent to flipping the signs of the Chern-Simons levels, $k_a\to-k_a$.  In the Fermi gas analysis of the necklace quivers, this corresponds to taking $c_i\to-c_i$, or equivalently $q\to-q$ in the Fermi gas Hamiltonian, (\ref{eq:HWlo}).  This essentially transforms the polygonal Fermi surface into its mirror image, but also flips the sign of $i\hbar$ in the quantum corrections.  Actually, it is easy to adjust the above calculation of the area of the Fermi surface by triangulating the polygon in the opposite sense.  In particular, instead of taking $c_{\hat s}$ to be the next largest parameter following $c_s$, we now take it to be the next smaller parameter.  At the same time, we have to take $\epsilon\to-\epsilon$ in the vertex shifts $f_s$ in (\ref{eq:fs}).  The end result is that $f_s\to-f_s$ under parity, and hence $\beta$ flips sign in (\ref{eq:coefs}), while $\alpha$, $\gamma$ and $\delta$ are unchanged.  This demonstrates explicitly that the partition function, (\ref{eq:Zairy2}), is mapped into its complex conjugate under parity.

Of course, for parity conserving quivers, the partition function should be real.  In particular, this indicates that $D=\beta/2\pi$ should vanish when parity is unbroken.  This is clearly true in the two-node (ABJM) case, and can also be directly verified for parity conserving four-node quivers.  However, we have been unsuccessful in demonstrating this explicitly for generic parity conserving quivers.  Note, of course, that all odd node quivers are necessarily parity violating as there will always be a mismatch between positive and negative Chern-Simons levels.

Finally, note that the Fermi gas approach is naturally applied in the M-theory limit where the overall Chern-Simons level $k=\hbar/2\pi$ is held fixed while $N$ is taken to infinity.  However, the free energy, (\ref{eq:Fexp}), which is obtained by expanding the Airy function can be reorganized in terms of a 't~Hooft expansion
\begin{equation}
    F=N^2F_0(\lambda)+F_1(\lambda)+\cdots,
\end{equation}
where $\lambda=N/k$.  The genus $g$ free energies, $F_g(\lambda)$, can be obtained in the large $\lambda$ limit by comparison with (\ref{eq:Fexp}).  In particular, we have
\begin{align}
    F_0(\lambda)&=-\fft23\mathcal C^{-1/2}\lambda^{-1/2}+i\fft{D}{2\mathcal C}\lambda^{-1}+\left(\mathcal B_2+\fft{D^2}{4\mathcal C}\right)\mathcal C^{-1/2}\lambda^{-3/2}+\mathcal O(\lambda^{-2}),\nn\\
    F_1(\lambda)&=\mathcal B_1\mathcal C^{-1/2}\lambda^{1/2}+\mathcal O(\lambda^0),
\end{align}
where
\begin{equation}
    \mathcal C=Ck=\fft\alpha{4\pi^2},\qquad\mathcal B_1=\left(B\big|_{\mathcal O(\hbar^{-1})}\right)k=\fft\delta{4\pi^2}+\fft\alpha{12},\qquad\mathcal B_2=\left(B\big|_{\mathcal O(\hbar)}\right)k^{-1}=\gamma.
\end{equation}
Note that the $B$ coefficient in (\ref{eq:BCD}) contributes at both genus zero and one.  Here the signature of parity violation shows up first at genus zero and is easily seen numerically.  Additional numerical investigations in the 't~Hooft limit suggest that both the $\mathcal O(\lambda^{-2})$ term in $F_0(\lambda)$ and the $\mathcal O(\lambda^0)$ term in $F_1(\lambda)$ are complex, indicating that $\tilde A(k)$ is indeed complex and takes the form
\begin{equation}
    \tilde A(k)=\tilde a_0k^2+\tilde a_1+\cdots=\tilde a_0N^2\lambda^{-2}+\tilde a_1N^0+\cdots,
\end{equation}
in the large $k$ limit.

In principle, it would be desirable to understand the sphere partition function, or equivalently the free energy $F(N,k)$, for arbitrary values of $N$ and $k$.  However, we do not expect that it would have a simple form in general.  Instead, we can study its behavior in various limits.  For holographic theories, it is generally sufficient to explore the large-$N$ limit.  However, even in this limit, there is the overall Chern-Simons level $k$ to consider.  The Fermi gas picture is most directly applied at fixed $k$, but involves the $N$ independent coefficient $\tilde A(k)$.  In the ABJM case, it is related to constant maps of the topological string \cite{Hanada:2012si}, but it is not clear to us whether such a connection persists for necklace quivers without parity invariance.  Additional, and somewhat complementary, information can be obtained in the 't~Hooft limit.  For example, the genus zero free energy $F_0(\lambda)$ can be obtained directly from the planar resolvent.  Although most investigations have focused on two node quivers, matrix resolvent techniques have been applied to more general necklace and linear quivers \cite{Suyama:2013fua,Suyama:2019dzv}.  Of course, for a complete picture to emerge, there are not only higher genus but also non-perturbative in $N$ corrections that need to be taken into account.  Our demonstration that the Airy function behavior persists in a large class of theories regardless of parity invariance gives additional emphasis to the universal applicability of the Fermi gas picture and strengthens the foundation for future investigations of the subleading structure of such $\mathcal N=3$ Chern-Simons-matter models.

\section*{Acknowledgments}

We gratefully acknowledge discussions with Junho Hong on computing subleading corrections in both Fermi gas and saddle point approaches. This work was supported in part by the U.S. Department of Energy under grant DE-SC0007859.


\bibliographystyle{JHEP}
\bibliography{cite.bib}

\end{document}